\documentclass[12pt,letterpaper]{JHEP3}

\usepackage{amscd,amsmath,amssymb,amsfonts,xspace,mathrsfs,amsthm}
\usepackage{color}
\let\normalcolor\relax
\usepackage{bbold}
\usepackage{latexsym}
\usepackage{graphicx}
\usepackage{dsfont}
\usepackage{color}
\usepackage{longtable}

  \newtheorem{conj}{Conjecture}

\numberwithin{equation}{section}

\def\lfig#1#2#3#4#5{
\begin{figure}[t]
 \centerline{\includegraphics[width=#3]{#2}}
 \vspace{#5}
  \caption{#1 \label{#4}}
 \end{figure}
}

\def\diag{{\rm diag}}

\def\Ch{{\rm Ch}}

\def\Sym{\,{\rm Sym}\, }

\def\Im{\,{\rm Im}\,}

\def\({\left(}
\def\){\right)}
\def\[{\left[}
\def\]{\right]}
\def\<{\left\langle}
\def\>{\right\rangle}
\def\hf{{1\over 2}}

\newcommand{\de}{\mathrm{d}}

\newcommand{\I}{\mathrm{i}}

\newcommand{\cL}{\mathcal{L}}
\newcommand{\cD}{\mathcal{D}}

\newcommand{\p}{\partial}

\newcommand{\cM}{\mathcal{M}}

\newcommand{\cN}{\mathcal{N}}

\newcommand{\cJ}{\mathcal{J}}

\DeclareSymbolFont{AMSa}{U}{msa}{m}{n}
\DeclareSymbolFont{AMSb}{U}{msb}{m}{n}
\DeclareMathSymbol{\fieldR}{\mathalpha}{AMSb}{"52}

\newcommand{\Fb}{{\mathbb F}}
\newcommand{\Bb}{{\mathbb B}}

\newcommand{\cI}{\mathcal{I}}
\newcommand{\cO}{\mathcal{O}}

\newcommand{\nn}{\nonumber}

\newcommand{\eps}{\epsilon}

\newcommand{\IT}{\mathds{T}}
\newcommand{\IR}{\mathds{R}}

\newcommand{\IC}{\mathds{C}}
\newcommand{\IZ}{\mathds{Z}}

\newcommand{\IH}{\mathds{H}}

\newcommand{\IP}{\mathds{P}}

\newcommand{\sgn}{\mbox{\rm sgn}}

\newcommand{\q}{\mbox{q}}

\def\bea{\begin{eqnarray}}
\def\eea{\end{eqnarray}}
\def\be{\begin{equation}}
\def\ee{\end{equation}}
\def\ba{\begin{align}}
\def\ea{\end{align}}
\def\bse{\begin{subequations}}
\def\ese{\end{subequations}}

\def\ba{\bar a}

\def\btau{\bar \tau}

\def\chS{\check S}

\def\cij#1{c}
\def\ci#1{c}

\def\XXint#1#2#3{{\setbox0=\hbox{$#1{#2#3}{\int}$}
\vcenter{\hbox{$#2#3$}}\kern-.5\wd0}}

\def\gamD#1{\tilde\gamma}

\def\CY{\mathfrak{Y}}

\def\cl0{\tilde c_0}

\newcommand{\bpt}{{\scriptstyle\boldsymbol{*}}}
\newcommand{\bfw}{{\boldsymbol w}}

\newcommand{\bfb}{{\boldsymbol b}}

\newcommand{\bfv}{{\boldsymbol v}}

\newcommand{\bfq}{{\boldsymbol q}}

\newcommand{\bfx}{{\boldsymbol x}}
\newcommand{\bfy}{{\boldsymbol y}}

\newcommand{\bfmu}{{\boldsymbol \mu}}
\newcommand{\bfnu}{{\boldsymbol \nu}}

\def\sE{\mathscr{E}}
\def\Zv{\mathscr{Z}}

\def\Bv{\mathscr{B}}

\def\Zv{\mathscr{Z}}

\def\Rv{\mathscr{R}}

\def\bOm{\bar\Omega}

\def\whh{\widehat h}

\def\whPhi{\widehat\Phi}

\def\sEp{\sE^{(+)}}
\def\sEf{\sE^{(0)}}

\def\Lat{\mathbf{\Lambda}}

\def\ptt{\mathscr{N}}
\def\bfptt{\boldsymbol{\mathfrak{p}}}

\def\cl{c^{(\ell)}}

\def\vu{\mathbb{u}}

\def\ve{\mathbb{e}}

\def\cLam{\check\Lambda}
\def\chh{\check h}
\def\chc{\check c}
\def\chw{\check w}
\def\chm{\check m}
\def\chq{\check q}

\def\chJ{\check J}
\def\chQ{\check Q}
\def\chC{\check C}
\def\whchh{\lefteqn{\widehat\chh}\phantom{h}}

\def\cmu{\check\mu}
\def\cgam{\check\gamma}
\def\gama{\hat\gamma}

\def\sN{M}
\def\y{y}

\def\nv{v_0}

\def\under#1#2{\mathop{#1}\limits_{#2}}

\def\bcup{\mathop{\Large\vphantom{A_a}\mbox{$\cup$}}}
\def\bplus{\mathop{\Large\vphantom{A_a}\mbox{$\oplus$}}}

\title{Rank $N$ Vafa-Witten invariants, modularity and blow-up}

\author{Sergei Alexandrov
\\
{\it
Laboratoire Charles Coulomb (L2C), Universit\'e de Montpellier,
CNRS, F-34095, Montpellier, France}\\

\vspace*{2mm} {\tt e-mail:
\email{sergey.alexandrov@umontpellier.fr}
}

\vspace*{-3mm}

}

\abstract{We derive explicit expressions for the generating functions of refined Vafa-Witten invariants $\Omega(\gamma,y)$
of $\mathds{P}^2$ of arbitrary rank $N$ and for their non-holomorphic modular completions.
In the course of derivation we also provide:
i) a generalization of the recently found generating functions of $\Omega(\gamma,y)$ and their completions
for Hirzebruch and del Pezzo surfaces in the canonical chamber of the moduli space to a generic chamber;
ii) a version of the blow-up formula expressed directly in terms of these generating functions
and its reformulation in a manifestly modular form.
}

\begin{document}

\section{Introduction}
\label{sec-intro}

The topologically twisted $\cN=4$ super Yang-Mills \cite{Vafa:1994tf} on a complex surface $S$, also known as Vafa-Witten (VW) theory,
appears to be at the heart of intersection of various theories, problems, approaches and conjectures in
mathematics and physics. On one hand, its partition function captures important topological information
about the surface $S$ encoded in the so called Vafa-Witten invariants, the Euler
numbers of the moduli spaces of semi-stable sheaves on $S$, which are the same as the moduli spaces of instantons
in the topologically twisted gauge theory.
The evaluation of these invariants is an important problem that has many links
to other mathematical subjects. In particular, it turns out that the generating functions of VW invariants exhibit non-trivial
modular properties and provide examples of either modular or mock modular forms \cite{Zwegers-thesis,MR2605321}.
This fact makes them a beautiful playground for number theory that can be used as a source for new ideas and identities.

On the other hand, the same invariants can be interpreted as BPS indices in supersymmetric gauge theory,
while their modular properties are understood as a consequence of S-duality \cite{Montonen:1977sn,Vafa:1994tf}.
Moreover, they can be related to generalized Donaldson-Thomas (DT) invariants of non-compact Calabi-Yau (CY) threefolds
because VW theory appears as an effective theory of M5-branes wrapped on $S\times T^2$ and reduced along the torus
where $S$ is a divisor of a CY.
Regarding these non-compact manifolds as local limits of compact CYs establishes further relations
with physics of BPS black holes, instantons and topological strings (see e.g. \cite{Minahan:1998vr}).

The connection to string theory outlined above furnishes us with a plethora of methods based on dualities and
other physical considerations that can bring new insights into the problem of evaluation of VW invariants
and understanding their modular properties.
In particular, recently an important progress has been achieved due to a complete
characterization of a modular anomaly of the generating functions of certain BPS indices in
string theory compactifications with $N=2$ supersymmetry \cite{Alexandrov:2018lgp,Alexandrov:2019rth}.
More precisely, reducing from M-theory to type IIA string, the above M5-brane system gives rise to a bound state
of D4-D2-D0-branes wrapped on even dimensional cycles of a compact Calabi-Yau threefold $\CY$, which at strong coupling
corresponds to a BPS black hole with a set of electro-magnetic charges. Hence, it makes sense to consider
a generating function of black hole degeneracies (obtained by summing over D0-brane charge),
whose mathematical interpretation is the generating function of the generalized DT invariants
counting semi-stable coherent sheaves supported on a divisor $\cD\subset\CY$. Whereas for $\cD$ {\it irreducible}
the generating function, evaluated in the so-called attractor chamber of the moduli space,
is known to be a modular form under S-duality group $SL(2,\IZ)$
\cite{Maldacena:1997de,Gaiotto:2005gf,deBoer:2006vg},
for {\it reducible} divisors where $\cD=\sum_{i=1}^n\cD_i$ it exhibits a modular anomaly \cite{Alexandrov:2016tnf}.
Remarkably, the anomaly turns out to be of a very special type and can be completely characterized by constructing
modular completions of the generating functions, i.e. their non-holomorphic deformations which
do transform as modular forms. Using duality constraints of string theory\footnote{These constraints require
the moduli space of compactified theory to carry an isometric action of $SL(2,\IZ)$.
At the same time, the metric on this moduli space receives D-instanton corrections weighted by DT invariants.
The completion was found by combining the isometry condition with the explicit description of D-instantons
in the twistor formalism \cite{Alexandrov:2008gh,Alexandrov:2009zh,Alexandrov:2012bu,Alexandrov:2012au}
(see \cite{Alexandrov:2011va,Alexandrov:2013yva} for reviews).},
in \cite{Alexandrov:2018lgp} a general formula for such completion in terms of the original holomorphic generating functions
has been found for a generic divisor. Its form implies that the generating functions are actually
vector valued {\it higher depth mock} modular forms, objects generalizing the notion of the usual mock modularity
\cite{bringmann2017higher}.

Then in \cite{Alexandrov:2019rth} this construction has been upgraded to include a refinement parameter $y$.
Its logarithm $z=\frac{\log y}{2\pi\I}$ transforms as an elliptic parameter so that the refined generating functions
turn out to be {\it higher depth mock Jacobi} forms.
Amazingly, the refined construction is in fact much simpler than the unrefined one and thus represents a natural framework
for further developments. In particular, taking the local limit on $\CY$ given by the canonical bundle
over a complex surface $S$ with $b_2^+(S)=1$ and $b_1(S)=0$, it has been used to get a formula for the modular completion
$\whh_{N,\mu}(\tau,z)$ of the generating functions $h_{N,\mu}(\tau,z)$
of refined VW invariants of $S$ with $U(N)$ gauge group \cite{Alexandrov:2019rth}.
The point is that for this class of surfaces such generating functions are known to have a modular anomaly \cite{Vafa:1994tf},
which can be traced back to the existence of non-holomorphic contributions to the partition function
of VW theory generated by Q-exact terms and coming from boundaries of the moduli space \cite{Dabholkar:2020fde}.
They ensure that the partition function is a true modular form given by the non-holomorphic
completion of the holomorphic generating functions.

Note that the formula of \cite{Alexandrov:2019rth} only expresses $\whh_{N,\mu}$ in terms of $h_{N_i,\mu_i}$ with $N_i\le N$,
which remained unknown at that stage. Nevertheless, it allows to trade the modular anomaly of the
generating functions for the holomorphic anomaly of their completions.
Combined with the requirement to have a well-defined unrefined limit $z\to 0$,
it can then be used to actually find the generating functions themselves,
similarly to solution of the topological string on elliptic CY threefolds \cite{Huang:2015sta,Gu:2017ccq}.
This program has been realized for Hirzebruch and del Pezzo surfaces, $\Fb_m$ and $\Bb_m$, in \cite{Alexandrov:2020bwg}
which produced a closed formula for both $h_{N,\mu}$ and $\whh_{N,\mu}$ for {\it all} ranks $N$. (For earlier results
on VW invariants of these surfaces, see
\cite{Yoshioka:1995,yoshioka1999euler,Manschot:2011dj,Manschot:2011ym,Klemm:2012sx,Beaujard:2020sgs}.)

This result however has two omissions.
First, as indicated above, the generating functions $h_{N,\mu}$ count the invariants only in a special chamber of the moduli space,
parametrized by a choice of polarization of $S$ provided by the K\"ahler form $J\in H^2(S,\IR)$.
This chamber corresponds to the attractor chamber of the CY geometry and,
in the terminology of \cite{Beaujard:2020sgs}, is called the {\it canonical} chamber, specified by $J=-K_S=c_1(S)$
where $K_S$ is the canonical class and $c_1(S)$ is the first Chern class of the surface.
In other chambers the invariants may have different values due to the wall-crossing phenomenon.
Although there are well established wall-crossing formulae which allow to compute the change of invariants from one chamber to another
\cite{ks,Joyce:2008pc,Joyce:2009xv}, it is certainly desirable to have an explicit formula for the generating functions
in arbitrary chamber.

Second, the construction of \cite{Alexandrov:2020bwg} relied on the existence of a null vector belonging to the integer homology lattice
$\Lambda_S=H_2(S,\IZ)$. Clearly, such vector does not exist in the case $b_2(S)=1$ corresponding to $S=\IP^2$.
Thus, this case remained uncaptured by the that construction.

In fact, the VW invariants of $\IP^2$ are determined by those of $\Fb_1$ by means of the
blow-up formula \cite{Yoshioka:1996,0961.14022,Li:1998nv}.
However this formula does not apply directly to $h_{N,\mu}$. Instead, one has to go to a different chamber
of the moduli space and pass through the so-called stack invariants \cite{joyce2008configurations}.
In principle, this is not a big deal and this procedure has been realized for any $N$ in \cite{Manschot:2014cca},
generalizing previous results in \cite{Yoshioka:1994,Manschot:2010nc,Manschot:2011ym} and expressing the generating functions
as combinations of generalized Appell functions.
However, the corresponding result for the completion, which is important not only as a characterization
of the modular properties of the  VW invariants, but also as a prediction for the partition function
of the physical theory, was missing so far. Of course, it can be obtained by either substituting
the generating functions of \cite{Manschot:2014cca} into the formula for the completion of \cite{Alexandrov:2019rth},
or by applying the general procedure developed in \cite{Alexandrov:2016enp} for constructing modular completions of
indefinite theta series and generalized Appell functions. But the former approach leads to expressions
with obscure modular properties, whereas the latter is very cumbersome and has been realized so far only for $N=3$
\cite{Manschot:2017xcr}.
On the other hand, the construction of \cite{Alexandrov:2020bwg} is designed to produce manifestly modular expressions
for completions, so that it is natural to try whether it can encompass the case of $\IP^2$ as well.
Besides, this might be useful keeping in mind its possible extension to finding DT invariants of compact CY threefolds
where the existence of a null vector belonging to the lattice is not guarantied either.

In this paper we fill in the above two gaps.
First, we conjecture (see eq. \eqref{genJhN}) an explicit expression for the generating functions of refined VW invariants of
Hirzebruch and del Pezzo surfaces for a two-parameter family of polarizations $J$
(which exhaust all possible $J$ in the case of $\Fb_m$).
Next, we show how one can avoid the obstruction of the absence of a null vector in $\Lambda_S$:
to this end, the formula for the modular completion $\whh_{N,\mu}$ can be rewritten
in terms of a set of functions $\chh_{N,\cmu}$ defined by an `extended lattice' $\cLam_S=\Lambda_S\oplus \IZ$
that can already possess such a vector.
Then the construction of \cite{Alexandrov:2020bwg} allows to find both $\chh_{N,\cmu}$ and their completions $\whchh_{N,\cmu}$,
while $h_{N,\mu}$ and $\whh_{N,\mu}$ follow from them in a trivial way.

It turns out that a particular case of this construction is nothing else but a variant of the blow-up formula.
However, it differs from the standard formula in that it is written directly in terms of the generating functions of VW invariants and
does not require the use of stack invariants. It can also be written in a way which is manifestly modular with respect
to the full duality group $SL(2,\IZ)$ (see eq. \eqref{compl-blowup}),
whereas usually only invariance under its congruence subgroup is apparent.
Thus, in a sense, our derivation provides a proof of consistency of the blow-up formula with modularity
for any rank $N$.

Specializing the above results for the case of $\IP^2$ allows to obtain explicit expressions for both
$h^{\,\IP^2}_{N,\mu}$ and $\whh^{\,\IP^2}_{N,\mu}$. The formula \eqref{whhP2} for the modular completion
is the main new result of this work.

The organization of the paper is as follows. In the next section we review the results of \cite{Alexandrov:2020bwg}
for $S=\Fb_m$ and $\Bb_m$ obtained in the canonical chamber of the moduli space and extend them to other chambers.
In section \ref{sec-latext} we present our construction based on the extended lattice, that allows to
find generating functions in the absence of a null vector in $\Lambda_S$, and relate it to the blow-up formula.
Then in section \ref{sec-P2} we apply it to the case of $\IP^2$. Section \ref{sec-concl} presents our conclusions.
Appendices contain useful details on theta functions, stack invariants, the blow-up formula and generalized error functions.

\section{Generating functions of refined Vafa-Witten invariants}
\label{sec-compl}

In this section we define the refined VW invariants, their generating functions and
discuss them for Hirzebruch and del Pezzo surfaces. The basic definitions are collected in $\S$\ref{subsec-VWinv}.
Then in $\S$\ref{subsec-results} we present the formula for $h_{N,\mu}$ in the canonical chamber found in \cite{Alexandrov:2020bwg}.
In $\S$\ref{subsec-gencham} it is generalized to a two-parameter family of polarizations.
Finally, in $\S$\ref{subsec-modcompl} we provide explicit results for the modular completions $\whh_{N,\mu}$.

\subsection{Refined VW invariants}
\label{subsec-VWinv}

Vafa-Witten invariants of a surface $S$ are defined as Euler numbers of the moduli spaces $\cM_{\gamma,J}$
of semi-stable coherent sheaves on $S$ where $\gamma=(N,\mu,\hf\mu^2-n)$ is the Chern character of the sheaf and
$J$ is a polarization entering the stability condition.\footnote{We refer to \cite{Beaujard:2020sgs}
and references therein for a more rigorous and detailed exposition.}
From physics point of view, $N$ is the rank of the gauge group $U(N)$, $J$ is the K\"ahler form,
whereas $\mu=-c_1(F)\in \Lambda_S$ and $n=\int_S c_2(F)\in \IZ$ are Chern classes of the gauge bundle.
The moduli space $\cM_{\gamma,J}$ parametrizes solutions of hermitian Yang-Mills equations
and depends on $J$ through the self-duality condition on the field strength $F$.
An important property, corresponding to the spectral flow symmetry,
is that the moduli space does not change upon tensoring $F$ with a line bundle $\cL$.
This leads to the shift $\mu\to \mu-Nc_1(\cL)$, but leaves invariant the rank $N$ and
the Bogomolov discriminant
\be
\Delta(F):= \frac{1}{N} \left( n - \frac{N-1}{2N} \mu^2 \right),
\label{Bogom}
\ee
where $\mu^2\equiv \int_S\mu^2$.
Due to this, the parameter $\mu$ can be restricted to $\Lambda_S/N\Lambda_S$.

In this paper we will be interested in the refined VW invariants, which are defined
by Poincar\'e polynomials of $\cM_{\gamma,J}$. More precisely, they are given by
\be
\Omega_J(\gamma,y)=\frac{\sum_{p=0}^{d} y^{2p-d_{\IC}(\cM_{\gamma,J})}\, b_p(\cM_{\gamma,J})}{y-y^{-1}}\, ,
\label{intOm}
\ee
where $y=e^{2\pi\I z}$ is the refinement parameter, $d_{\IC}(\cM)$ is the complex dimension of $\cM$
and $b_p(\cM)$ is its Betti number.
The parameter $z$ is taken to be complex $z=\alpha-\tau\beta$ with both $\alpha$ and $\beta$ non-vanishing.
To exhibit modular properties of the refined invariants, one has to work actually in terms of their rational counterparts
\cite{Manschot:2010xp,Manschot:2017xcr}
\be
\bOm_J(\gamma,y) =  \sum_{m|\gamma} \frac{1}{m}\, \Omega_J(\gamma/m, - (-y)^{m}).
\label{defcref}
\ee
It is these invariants that we use to define the generating functions\footnote{Sometimes we will drop the upper index
indicating the surface when it does not cause a confusion.}
\be
\label{defhVWref}
h^S_{N,\mu,J}(\tau,z) =
\sum_{n\geq 0}
\bOm_J(\gamma,y)\,
\q^{N\(\Delta(F) - \tfrac{\chi(S)}{24}\)},
\ee
where as usual $\q=e^{2\pi\I\tau}$.
As follows from \eqref{intOm}, this function has single poles at $z=0$ and $z=\hf$ with the residues
given by the generating function of the unrefined VW invariants.

For $N=1$, there is no dependence on $\mu$ and $J$, and
the generating function is known for any $S$ \cite{Gottsche:1990}.
For surfaces with $b_1(S)=0$, it is given by
\be
\label{h10anySref}
h^S_{1,0}(\tau,z) = \frac{\I}{\theta_1(\tau,2z)\, \eta(\tau)^{b_2(S)-1}},
\ee
where $\theta_1(\tau,z)$ is the Jacobi theta function and $\eta(\tau)$ is the Dedekind eta function.
For higher ranks one obtains already a vector valued function with components labelled by $\mu\in\Lambda_S/N\Lambda_S$.
Regarding dependence on $J$, when $b_2^+(S)>1$ it was found to be absent, but
for $b_2^+(S)=1$ and $b_2(S)>1$ it does affect the generating functions.
Of course, this dependence is piecewise constant and can be captured by the standard wall-crossing formulae
\cite{ks,Joyce:2008pc,Joyce:2009xv}.
In the following, we will be interested in surfaces with $b_2^+(S)=1$ and $b_1(S)=0$
which include, in particular, Hirzebruch $\Fb_m$, del Pezzo $\Bb_m$ and $\IP^2$.
Thus, for all of them except $\IP^2$ there is a non-trivial wall-crossing.

\subsection{Generating functions for $\Fb_m$ and $\Bb_m$: canonical chamber}
\label{subsec-results}

Among all possible choices of polarization $J$ there is one special $J=-K_S$ which corresponds to the canonical chamber in
the moduli space. On one hand, this is the chamber with the richest BPS spectrum \cite{Beaujard:2020sgs}.
On the other hand, after uplifting to  a CY threefold, it corresponds to the attractor chamber of the CY moduli space where
the moduli are fixed by the charges and one expects that no bound states are present \cite{Alexandrov:2018iao}
(except the so-called scaling solutions \cite{Denef:2007vg,Bena:2012hf}).
This is the chamber where the construction of \cite{Alexandrov:2018lgp,Alexandrov:2019rth} applies,
and in \cite{Alexandrov:2020bwg} it was used to get an explicit formula for the generating functions $h^S_{N,\mu}:=h^S_{N,\mu,-K_S}$
for Hirzebruch and del Pezzo surfaces, which we now review.

First, let us recall a few basic facts about these surfaces.
The Hirzebruch surface $\Fb_m$ is a projectivization of the $\cO(m)\oplus \cO(0)$ bundle over $\IP^1$.
It has $b_2(S)=2$ and in the basis given by the curves corresponding to the fiber $[f]$ and the section of the bundle $[s]$,
the intersection matrix and the first Chern class are the following
\be
C_{\alpha\beta} = \begin{pmatrix} 0 & 1 \\ 1 & -m \end{pmatrix},
\qquad
c_1(\Fb_m) = (m+2)[f]+2[s].
\label{dataFk}
\ee
The del Pezzo surface $\Bb_m$ is the blow-up of $\IP^2$ over $m$ generic points.
It has $b_2(S)=m+1$ and a basis of $\Lambda_{\Bb_m}$ is given by the hyperplane class of $\IP^2$ and
the exceptional divisors of the blow-up denoted, respectively, by $D_1$ and $D_2,\dots,D_{m+1}$.
In this basis the intersection matrix and the first Chern class are given by
\be
C_{\alpha\beta}= \diag (1,-1,\dots,-1),
\qquad
c_1(\Bb_m) = 3D_1 -\sum_{\alpha=2}^{m+1}D_\alpha.
\label{dataBk}
\ee
A few comments are in order:
\begin{itemize}
\item
For all above surfaces as well as for $\IP^2$, the signature of the intersection matrix is $(1,b_2-1)$ and  
\be
c_1^2(S)=10-b_2(S).
\label{c1b2}
\ee
\item
In fact, $\Fb_1=\Bb_1$, i.e. it is the blow-up of $\IP^2$ at one point,
and by changing the basis to
\be
D_1=[f]+[s],
\qquad
D_2=[s],
\label{baseF1}
\ee
one brings $C_{\alpha\beta}$ and $c_1(\Fb_1)$ to the form \eqref{dataBk} with $m=1$.

\item
The crucial role in the construction of the generating functions of VW invariants is played by a null vector $\nv\in\Lambda_S$.
For $S=\Fb_m$ and $\Bb_m$, this vector must be chosen as follows \cite{Alexandrov:2020bwg}
\be
\nv(\Fb_m)=[f],
\qquad
\nv(\Bb_m)=D_1-D_2.
\label{nullv}
\ee
\end{itemize}

Next, one has to introduce several notations.
Let $\gama=( N, q)$ where $q\in \Lambda_S+\frac{N}{2} K_S$ and it is further decomposed
into the part spanning $N\Lambda_S$ and the residue class $\mu\in\Lambda_S/N\Lambda_S$,
that reflects the spectral flow symmetry.
In a basis $D_\alpha$, $\alpha=1,\dots, b_2(S)$, of $H_2(S,\IZ)$ this decomposition is given by
\be
\label{quant-q}
q_\alpha =  N \, C_{\alpha\beta} \epsilon^{\beta} + \mu_\alpha- \frac{N}{2}\, C_{\alpha\beta} c_1^\beta,
\qquad
\eps^\alpha\in \IZ,
\ee
where $C_{\alpha\beta}=D_\alpha\cap D_\beta$ and $c_1^\alpha$ are the components 
of the first Chern class $c_1(S)=-K_S=c_1^\alpha D_\alpha$ of the surface.
Then we introduce a combination of theta series\footnote{In \cite{Alexandrov:2020bwg} the same notation was used
for the normalized combinations differing from \eqref{combTh} by the factor of $(h_{1,0}^S)^N$.}
\be
\Theta^S_{N,\mu}(\tau,z;\{\Phi_n\})=\sum_{n=1}^\infty\frac{1}{2^{n-1}}
\sum_{\sum_{i=1}^n \gama_i=\gama}
\Phi_n(\{\gama_i\})
\, \q^{\hf Q_n(\{\gama_i\})}
\, \y^{\sum_{i<j}\gamma_{ij}(c_1(S))}\prod_{i=1}^n H^S_{N_i,\mu_i}(\tau,z),
\label{combTh}
\ee
where the sum goes over all decompositions of the charge $\gama=(N,\mu- \frac{N}{2}c_1)$,
i.e. with the spectral flow parameter set to zero, and the charges
$q_{i}$ are quantized as in \eqref{quant-q} with $N$ replaced by $N_i$.
The other notations used in \eqref{combTh} are the following:
\begin{itemize}
\item
the quadratic form $Q_n$ given by
\be
Q_n(\{\gama_i\})= \frac{1}{N}\, q^2-\sum_{i=1}^n \frac{1}{N_i} q_{i}^2
=-\sum_{i<j}\frac{(N_i q_j - N_j q_i)^2}{NN_iN_j}\, ,
\label{defQlr}
\ee
where $q^2=C^{\alpha\beta}q_\alpha q_\beta$ and $C^{\alpha\beta}$ is the inverse of $C_{\alpha\beta}$;
for the charges satisfying $\sum_i q_i=q$ with $q$ fixed, the signature of $Q_n$ is $(n-1,(n-1)(b_2-1))$;

\item
anti-symmetrized Dirac product of charges depending on a vector $v\in \Lambda_S$
\be
\label{gam12}
\gamma_{ij}(v)=
v^{\alpha} (N_i q_{j,\alpha} -N_j q_{i,\alpha});
\ee

\item
the generating function $H^S_{N,\mu}$ of stack invariants evaluated at $J=\nv$, which is given in \eqref{gfHN},

\item
the kernels $\Phi_n$ depending on $n$ charges that determine the theta series and will be specified below.

\end{itemize}
Note that both \eqref{defQlr} and \eqref{gam12} are invariant under an overall shift of the spectral flow parameters $\eps_i^\alpha$.
The same is supposed to be true (and indeed will be) for the kernels $\Phi_n$.
Therefore, the r.h.s. of \eqref{combTh} is invariant under the spectral flow shift of $q$,
which explains why this symmetry is fixed in $\gama$.

The main result of \cite{Alexandrov:2020bwg} is that the generating functions of refined VW invariants in
the canonical chamber are expressed through the combinations \eqref{combTh}
\be
h^S_{N,\mu}=\Theta^S_{N,\mu}(\tau,z;\{\Phi_n(-K_S)\})
\label{whgN}
\ee
with the following kernels
\be
\Phi_n(\{\gama_i\};v)
=\sum_{\cJ\subseteq \Zv_{n-1}}\[e_{|\cJ|}\,\delta_\cJ(v)
\prod_{k\in \Zv_{n-1}\setminus \cJ}\Bigl(\sgn(\Gamma_k(v))-\sgn(\Bv_k)\Bigr)\],
\label{kerg}
\ee
where $\Zv_{n}=\{1,\dots,n\}$, $|\cJ|$ is the cardinality of the set,
$e_{m-1}$ is the $m$-th Taylor coefficient of ${\rm arctanh}$, namely
\be
e_m=\left\{ \begin{array}{cc}
0 & \mbox{ if }m \mbox{ is odd},
\\
\frac{1}{m+1}& \mbox{ if }m \mbox{ is even},
\end{array}\right.
\label{valek}
\ee
$\delta_\cJ(v)=\prod_{k\in \cJ}\delta_{\Gamma_k(v)}$ is the product of Kronecker deltas, and
\be
\begin{split}
\Gamma_{k}(v)=&\,\sum_{i=1}^k\sum_{j=k+1}^n \gamma_{ij}(v),
\\
\Bv_k=&\, \gamma_{k,k+1}(\nv)+\beta N_k N_{k+1}(N_k+N_{k+1})\, \nv\cdot c_1(S).
\end{split}
\label{defbk}
\ee
The structure of the kernel \eqref{kerg} is in fact very simple. Generically,
it is given by a product of differences of two sign functions.
This is a typical structure for a kernel of indefinite theta series ensuring their convergence \cite{Alexandrov:2016enp}.
In our case, one set of sign functions is determined by the null vector $\nv$ \eqref{nullv} and the second
by the vector $v$, which according to \eqref{whgN} is equal to the anti-canonical class $-K_S=c_1(S)$.
However, for some values of charges the arguments of the second set
may vanish.\footnote{For $\Bv_k$ this does not happen for generic $\beta$.} Such situations require a special attention.
The formula \eqref{kerg} is written using convention $\sgn(0)=0$, however it turns out that the construction
taking its roots in modularity requires, roughly speaking,
that a product of even number of sign functions with vanishing arguments is to be replaced by
the non-vanishing number $e_m$, i.e. $(\sgn(0))^m\to e_m$. This is the origin of the sum over subsets in \eqref{kerg}.

\subsection{Generating functions for $\Fb_m$ and $\Bb_m$: generic chamber}
\label{subsec-gencham}

Given the result \eqref{whgN}, it is easy to guess an expression for the generating functions
for a more general polarization $J$:

\begin{conj}
For $J\in\mbox{\rm Span}(c_1(S),\nv(S))^+$, where $\nv(S)$ is specified in \eqref{nullv}
and $+$ indicates the restriction to the K\"ahler cone, one has
\be
\fbox{$h^S_{N,\mu,J}=\Theta^S_{N,\mu}(\tau,z;\{\Phi_n(J)\})$}
\label{genJhN}
\ee
\label{conj-J}
\end{conj}

\vspace{-0.2cm}
Note that for $S=\Fb_m$ the restriction on $J$ is empty since $\mbox{\rm Span}(c_1(S),\nv)^+$ coincides with the full K\"ahler cone.
On the other hand, for $S=\Bb_m$ it restricts $J$ to belong to a two-dimensional plane in the $m+1$ dimensional moduli space.
This restriction is necessary because for more general $J$ the formula \eqref{genJhN} appears to not have 
a well-defined unrefined limit and to be inconsistent with
the formula for the completion \eqref{exp-whhr} given below (see footnote \ref{foot-J}).

We will not try to prove this conjecture. Instead, we provide several evidences in its favor:
\begin{itemize}
\item
We have checked numerically that it agrees with the formulae for $h^{\Fb_m}_{N,\mu,J_{k,\ell}}$, $N=2,3$,
where $J_{k,\ell}=k([s]+m[f])+\ell [f]$, $k,\ell\ge 0$, which were given in \cite{Manschot:2011ym},
provided one does not hit a wall of marginal stability what may happen for $J\cdot\mu =0\!\! \mod N$.

\item
In the canonical chamber $J=-K_S$ it reduces to the result \eqref{whgN}.

\item
For $J=\nv$ it reduces to the generating function of stack invariants $H^S_{N,\mu}$ \eqref{gfHN}
plus the so-called `zero mode contributions'
where some of the charges are fixed by the condition that at least one of $\Gamma_k(\nv)$ vanishes.
This is the expected structure since the zero mode contributions account for the difference between
stack and VW invariants.

\item
As we will se below, it nicely agrees with the blow up formula.

\end{itemize}

\subsection{Modular completion}
\label{subsec-modcompl}

As explained in the Introduction, the result \eqref{whgN} has been derived
by requiring a proper behavior under modular transformations
which act on the arguments of the generating functions in the following way
\be
\tau\to \frac{a\tau+b}{c\tau+d}\, ,
\qquad
z\to \frac{z}{c\tau+d}\, ,
\qquad
\(\begin{array}{cc}
a & b \\ c & d
\end{array}\)\in SL(2,\IZ).
\label{transf-tz}
\ee
Identifying $\tau$ with the complexified coupling constant of $\cN=4$ super-Yang-Mills, one recognizes
in \eqref{transf-tz} the standard S-duality which leaves this theory invariant. This suggests that
the partition function of VW theory transforms as a modular form. Furthermore, since the dependence on $\btau$
can be shown to be Q-exact \cite{Vafa:1994tf}, the standard arguments imply that the partition function is holomorphic in $\tau$
and is essentially captured (up to a simple theta series) by the generating function $h_{N,\mu}$.
The latter therefore is expected to behave as a (vector valued) Jacobi form \cite{MR781735}.
The transformation properties of such objects are shown in \eqref{Jacobi} and characterized by two numbers: weight and index.
In our case they are given by \cite{Alexandrov:2019rth}
\be
\label{indexconjVW}
w_S = -\frac12\, b_2(S),
\qquad
m_S(N) = -\frac16\, K_S^2 (N^3-N)-2N .
\ee

All these expectations turn out to be true for $h^S_{1,0}$ \eqref{h10anySref}.
However, for $N\ge 2$ and $b_2^+(S)=1$ they were found to fail \cite{Vafa:1994tf}.
In this case the non-holomorphic contributions from Q-exact terms can be shown to be non-vanishing
due to boundaries of the moduli space \cite{Dabholkar:2020fde}.
As a result, the full partition function is not holomorphic anymore and does not reduce to $h_{N,\mu}$,
while it is still believed to be modular.
This implies that $h_{N,\mu}$ has a modular anomaly, but at the same time it has a non-holomorphic modular completion
which then gives the partition function. An explicit expression for such completion is one of our main interests.

For $S=\Fb_m$ and $\Bb_m$, the modular completion was found in \cite{Alexandrov:2020bwg} in the canonical chamber,
but given Conjecture \ref{conj-J} it can easily be generalized to the class of polarizations described there.
Indeed, there is a general recipe \cite{Alexandrov:2016enp,Nazaroglu:2016lmr}
for constructing completions of indefinite theta series with kernels given by combinations of
sign functions, which is explained in appendix \ref{ap-generr}.
Applying it to the generating functions \eqref{genJhN},
one obtains\footnote{We assume that the modular group does not act on the polarization $J$.
In VW theory this is not true and it transforms as $J\to J/|c\tau+d|$. However, a rescaling of $J$ does not affect VW invariants.
Therefore, to take into account the transformation of $J$, it is enough to replace in all equations below $J$ by 
$J/\sqrt{\tau_2}$, which does stay invariant. We avoid from doing so 
explicitly because then the specialization to the canonical chamber would read $J=\sqrt{\tau_2}\,c_1$.}
\be
\fbox{$\whh^{\,S}_{N,\mu,J}(\tau,z)=\Theta^S_{N,\mu}(\tau,z;\{\whPhi_n(J)\})$}
\label{complFBJ}
\ee
where the kernels
\be
\whPhi_n(\{\gama_i\};J)=\sum_{\cJ\subseteq \Zv_{n-1}} \Phi_{|\cJ|}^E(\{ \bfv_{\ell}(J)\}_{\ell\in \cJ};\bfx)
\prod_{k\in \Zv_{n-1}\setminus \cJ}\Bigl(-\sgn(\Bv_k)\Bigr)
\label{kerhg}
\ee
are expressed through the generalized error functions $\Phi_n^E$ \eqref{generrPhiME}
with the vectors $\bfx$ and $\bfv_\ell(J)$ defined in \eqref{vectors}.

We also need another expression for the completion which expresses it through the holomorphic functions $h_{N_i,\mu_i}$
and holds more generally than for $S=\Fb_m$ and $\Bb_m$.
It has been derived in \cite{Alexandrov:2019rth} for $J=-K_S$ and was the key for deriving \eqref{whgN} in \cite{Alexandrov:2020bwg}.
Here, in accordance with the previous discussion, we generalize it to $J\in\mbox{\rm Span}(c_1,\nv)^+$
in which case it reads\footnote{Inspecting the derivation of \eqref{whgN} from \eqref{exp-whhr} done in \cite{Alexandrov:2020bwg}
for $J=c_1$, it is a easy to see that the only place where it depends on the polarization is the derivation
of the holomorphic modular factor given by the functions $H^S_{N,\mu}$. However, this derivation involves only charges
satisfying $\nv\cdot q_i=0$, while for such charges and $J\in\mbox{\rm Span}(c_1,\nv)^+$,
one has $J\cdot q_i=ac_1\cdot q_i$ with positive $a$. Since the coefficient $a$ is canceled in sign functions, this implies that
the derivation still goes through. This provides a justification for \eqref{exp-whhr} and at the same time demonstrates
the need of the restriction on $J$. \label{foot-J}}
\be
\fbox{$\displaystyle
\whh^{\,S}_{N,\mu,J}(\tau,z)= \sum_{n=1}^\infty\frac{1}{2^{n-1}}
\sum_{\sum_{i=1}^n \gama_i=\gama}
\Rv_n(\{\gama_i\},J)
\, \q^{\hf Q_n(\{\gama_i\})}
\, \y^{\sum_{i<j} \gamma_{ij}(c_1) }
\prod_{i=1}^n h^S_{N_i,\mu_i,J}(\tau,z)$}
\label{exp-whhr}
\ee
It involves the same objects which already appeared in \eqref{combTh}.
The only new one is the coefficient $\Rv_n$. We will not need its explicit expression and
therefore relegate its definition to appendix \ref{ap-Rn}.
The only relevant information for us is that it is expressed through the same generalized error functions
which appeared in \eqref{kerhg}.

\section{Modularity, lattice extension and blow-up}
\label{sec-latext}

In this section we would like to address the problem which arises when the lattice $\Lambda_S$ does not possess
a null vector, which plays the crucial role in the construction of \cite{Alexandrov:2020bwg} and appears explicitly
in the definition of kernels \eqref{kerg} and \eqref{kerhg}. An obvious example of such situation
is $S=\IP^2$ where the lattice $\Lambda_{\IP^2}$ is one-dimensional. But we will not restrict ourselves to this
particular case and proceed in full generality since this might find applications in attempts to extend the present approach
to calculation of DT invariants of compact CY threefolds. As we will see, it pays off because
it allows to rederive the blow-up formula, reviewed in appendix \ref{ap-stack},
from the formula for the modular completion \eqref{exp-whhr}
and moreover to give it a slightly new formulation.

\subsection{Generating functions from lattice extension}

Our starting point is the equation \eqref{exp-whhr} expressing the modular completion $\whh^S_{N,\mu,J}$
through the holomorphic generating functions $h^S_{N_i,\mu_i,J}$.
Let us multiply this equation by the `blow-up function'
$B_{N,\ell}(\tau,\kappa z)= \theta^{(N)}_{\ell}(\tau,\kappa z)/\eta^{N}(\tau)$ \eqref{defBNk} where $\kappa\in \IZ$.
The parameter $\kappa$ is introduced here due to the following reason.
On one hand, taking $\kappa=0$ seems to be the simplest possibility which allows to perform the manipulations below.
On the other hand, the choice $\kappa=1$ appears to be the most natural one from the geometric point of view
since it will be shown to produce the blow-up construction.
To keep both possibilities available, we prefer to keep $\kappa$ generic.

Now, in each term on the r.h.s. of the resulting equation we apply to the theta function $\theta^{(N)}_{\ell}$
the identity \eqref{decompThN} with $N_i$ equal to the components of the charges $\gama_i$.
Then $\eta^{-N}$ and all factors $\theta^{(N_i)}_{\ell_i}$ can be absorbed into a redefinition of the generating functions
and their completions
\be
\begin{split}
\chh^S_{N,\cmu,J}(\tau,z)=&\, B_{N,\ell}(\tau,\kappa z)\, h^S_{N,\mu,J}(\tau,z),
\\
\whchh^{\,S}_{N,\cmu,J}(\tau,z)=&\, B_{N,\ell}(\tau,\kappa z) \,\whh^{\,S}_{N,\mu,J}(\tau,z),
\end{split}
\label{defchh}
\ee
where $\cmu=(\mu,\ell)$. Given the modular properties of $B_{N,\ell}$, the new functions $\whchh^{\,S}_{N,\cmu,J}$
must behave as vector valued Jacobi forms with weight and index given by
\be
\label{indexconjVWch}
\chw_S = -\frac12\, (b_2(S)+1),
\qquad
\chm_S(N) = \frac16\, (N^3-N)(\kappa^2- K_S^2)-2N .
\ee
In this way one arrives at the following equation
\be
\whchh^{\,S}_{N,\cmu,J}= \sum_{n=1}^\infty\frac{1}{2^{n-1}}
\sum_{\sum_{i=1}^n \gama_i=\gama}
\Rv_n(\{\gama_i\},J)
\, \q^{\hf Q_n(\{\gama_i\})}
\, \y^{\sum_{i<j} \gamma_{ij}(c_1) }
\sum_{\ell_i=0}^{N_i-1}\theta^{(\vec N)}_{\ell,\vec\ell}(\tau,\kappa z)\prod_{i=1}^n\chh^S_{N_i,\cmu_i,J}.
\label{exp-whhr-ch}
\ee
The form of the theta series $\theta^{(\vec N)}_{\ell,\vec\ell}(\tau,\kappa z)$ \eqref{defthNll} and the relation
\eqref{sumgam} for the power of the refinement parameter suggest that the theta series can also be absorbed.
To this end, let us extend the lattice $\Lambda_S$ by adding to it the trivial integer lattice,
$\cLam_S=\Lambda_S\oplus \IZ$, so that the new bilinear form is given by
\be
\chC=\(\begin{array}{cc}
C & 0 \\  0 & -1
\end{array}\).
\label{extC}
\ee
Following this idea we also define $\cgam=(N,\chq)$ with $\chq\in N\cLam_S+\cmu-\frac{N}{2}\, \chc_1$
and $\chc_1=(c_1, -\kappa)$. Finally, we take the quadratic form $\chQ_n$ to be the same as in \eqref{defQlr}, but evaluated on
the extended lattice.
It is easy to check that, rewritten in terms of the extended lattice, eq. \eqref{exp-whhr-ch}
takes exactly the same form as the initial equation \eqref{exp-whhr}
\be
\whchh^{\,S}_{N,\cmu,J}(\tau,z)= \sum_{n=1}^\infty\frac{1}{2^{n-1}}
\sum_{\sum_{i=1}^n \cgam_i=\cgam}
\Rv_n(\{\cgam_i\},J)
\, \q^{\hf \chQ_n(\{\cgam_i\})}
\, \y^{\chc_1\cdot\sum_{i=1}^n \ptt_i \chq_i }
\prod_{i=1}^n \chh^S_{N_i,\cmu_i,J}(\tau,z).
\label{exp-whhr-c}
\ee
Note that we could replace in the coefficient $\Rv_n$ the charges $\gama_i$ by the new ones $\cgam_i$ because
its dependence on charges and the first Chern class $c_1$ is captured by the scalar products
\eqref{defGam} and the contraction with $J$ ensures that the additional components of $\cgam_i$ and $\chc_1$ do not contribute.

Thus, we have found that the functions $\chh^S_{N,\cmu,J}$ and their completions satisfy the same equations
as the generating functions of refined VW invariants of, may be fictional or may be actual, surface $\chS$
with $b_2(\chS)=b_2(S)+1$, bilinear form given by \eqref{extC} and the first Chern class $c_1(\chS)=\chc_1$.
This allows to conclude that
\be
\begin{split}
\chh^S_{N,\cmu,J}(\tau,z)=&\,  h^{\chS}_{N,\cmu,\iota(J)}(\tau,z),
\\
\whchh^{\,S}_{N,\cmu,J}(\tau,z)=&\, \whh^{\,\chS}_{N,\cmu,\iota(J)}(\tau,z),
\end{split}
\label{defchhchS}
\ee
where $\iota(J)=(J,0)$ is the embedding $\iota:\Lambda_S\hookrightarrow \Lambda_{\chS}=\cLam_S $.
The difference with respect to the original problem of finding $h^S_{N,\mu,J}$
and their completions is that the new lattice $\Lambda_{\chS}$ may have a null vector even if the old one $\Lambda_S$
did not have it. If $\Lambda_{\chS}$ still does not have a null vector, one can repeat the construction
by further extending the lattice until it has it.\footnote{Such situation may happen for non-unimodular lattices. For instance, if
$\Lambda_S=\IZ$ with quadratic form equal to $2k^2$, $k\in\Lambda_S$,
its extension does not have null vectors because the null vectors of quadratic form $\diag(2,-1)$ are irrational.
In this case one would have to extend $\Lambda_S$ twice.
See \cite{Alexandrov:2017qhn} for an example of using such construction for proving S-duality of
multiple D3-instantons in type IIB string theory on a CY threefold.}
Assuming that such vector does exist, one can apply the approach of \cite{Alexandrov:2020bwg} to get the generating functions and
their completions for $\chS$.\footnote{If $\chS$ is a fictional surface, the existence of a solution is not guaranteed!}
In particular, if $\chS$ coincides with one of Hirzebruch or del Pezzo surfaces, one could immediately
borrow results from \eqref{genJhN} and \eqref{complFBJ}. In any case, the generating functions and their completions for
the initial surface $S$ then follow from \eqref{defchh} and \eqref{defchhchS}, i.e.
\be
\begin{split}
h^S_{N,\mu,J}(\tau,z)=&\, \frac{h^{\chS}_{N,(\mu,\ell),\iota(J)}(\tau,z)}{B_{N,\ell}(\tau,\kappa z)}\,,
\qquad
\whh^{\,S}_{N,\mu,J}(\tau,z)=\frac{\whh^{\,\chS}_{N,(\mu,\ell),\iota(J)}(\tau,z)}{ B_{N,\ell}(\tau,\kappa z)} \,.
\end{split}
\label{relhch}
\ee
Note that the index $\ell$ on the r.h.s. is arbitrary. Therefore, the functions $h^{\chS}_{N,\cmu,\iota(J)}$
must satisfy the following integrability conditions
\be
B_{N,k}(\tau,\kappa z) \, h^{\chS}_{N,(\mu,\ell),\iota(J)}(\tau,z)=B_{N,\ell}(\tau,\kappa z) \, h^{\chS}_{N,(\mu,k),\iota(J)}(\tau,z)
\label{inegrcond}
\ee
to ensure the independence of the ratios \eqref{relhch} on this index.\footnote{Similar conditions
for $\whh^{\,\chS}_{N,\cmu,\iota(J)}$ will then automatically follow due to the modular properties of the blow-up functions.}
Of course, these conditions trivially hold for \eqref{defchh}, but the point is that the construction of the generating functions
of $\chS$ does not know {\it a priori} about them and they can be extremely non-trivial.
In fact, they can be viewed as consistency conditions of the whole construction.

There is however one important drawback of the relations \eqref{relhch}:
they make the modular properties of the completion $\whh^{\,S}_{N,\mu,J}$ unobvious.
It is unclear why a ratio of two vector valued modular functions should itself form
a vector representation of the modular group.
Fortunately, one can get an alternative representation with help of the identity \eqref{diagThN}.
Applying it to \eqref{defchh}, one obtains
\be
\begin{split}
h^S_{N,\mu,J}(\tau,z)=&\, \frac{\eta(\tau)^N}{\prod_{j=1}^N \theta_1(\tau,\kappa_j z)}
\sum_{\ell=0}^{N-1}\theta_{N,\ell}(\tau,\kappa' z)\,h^{\chS}_{N,(\mu,\ell),\iota(J)}(\tau,z),
\\
\whh^{\,S}_{N,\mu,J}(\tau,z)=&\,  \frac{\eta(\tau)^N}{\prod_{j=1}^N \theta_1(\tau,\kappa_j z)}
\sum_{\ell=0}^{N-1}\theta_{N,\ell}(\tau,\kappa' z)\,\whh^{\,\chS}_{N,(\mu,\ell),\iota(J)}(\tau,z),
\end{split}
\label{h-chh}
\ee
where $\kappa_j=\kappa'+(2j-N-1) \kappa$ and $\kappa'$ should be chosen such that none of $\kappa_j$ vanishes.
In this representation the numerator is a contraction of two modular vectors and denominator
involves only modular scalars so that the result is manifestly modular.
In particular, the modular properties of theta functions and
$\whh^{\,\chS}_{N,\cmu,\chJ}$ ensure that $\whh^{\,S}_{N,\mu,J}$
does transform as a Jacobi form of weight and index specified in \eqref{indexconjVW}.

\subsection{Relation to the blow-up formula}

The above construction depends on the integer valued parameter $\kappa$.
It affects the auxiliary surface $\chS$ through its first Chern class $\chc_1=(c_1,-\kappa)$
and appears in the relations \eqref{relhch} and \eqref{h-chh} between generating functions.
It is unlikely that the construction goes through and gives the same result for any $\kappa$.
So what should be the value of this parameter?

The simplest possibility is to set $\kappa=0$. This simplifies various equations including
\eqref{h-chh} where one can take $\kappa_j=\kappa'=1$. However, there is no guarantee that there exists
$\whh^{\,\chS}_{N,\cmu,\chJ}$ with all required properties.
For instance, if $S=\IP^2$, $\Fb_m$ or $\Bb_m$, the first Chern class of $\chS$ satisfies
\be
\chc_1^2=c_1^2(S)=10-b_2(S)=11-b_2(\chS),
\ee
and thus spoils \eqref{c1b2}, so that $\chS$ is probably fictional and a solution for generating functions may not exist.
And indeed, in the case $S=\IP^2$ one can show that it is impossible to find an analogue of the functions
$H^{\Fb_m}_{N,\mu}(\tau,z)$, which would have the same behavior near $z=0$ and $z=1/2$
but transform with index (see eq. \eqref{indexconjVWch})
$\chm_{\IP^2}(N)=-\frac12\, (3N^3+N)$ instead of $-\frac23\, (2N^3+N)$.

Another natural choice is $\kappa=1$. In this case the multiplication factors in the relations \eqref{relhch}
become identical to the standard blow-up functions, whereas the relations themselves are reminiscent the blow formula
\eqref{blowupf}. This is not an accident since a blow-up of $S$ gives a surface $\chS$ with exactly the same data
as one obtains for $\kappa=1$: the intersection matrix \eqref{extC} and the first Chern class $\chc_1=c_1-D_e$,
where $D_e$ is the exceptional divisor of the blow-up responsible for the factor $\IZ$ in $\Lambda_{\chS}=\Lambda_S\oplus\IZ$.
Thus, in this case our construction reproduces the standard blow-up and shows its consistency with
the modular properties of the completions.
In particular, choosing in \eqref{h-chh} $\kappa'=N$, one arrives at the following manifestly modular relation
equivalent to the blow-up formula
\be
\fbox{$\displaystyle\whh^{\,S}_{N,\mu,J}(\tau,z)=  \frac{\eta(\tau)^N}{\prod_{j=1}^N \theta_1(\tau,(2j-1) z)}
\sum_{\ell=0}^{N-1}\theta_{N,\ell}(\tau,N z)\,\whh^{\,\chS}_{N,\iota(\mu)+\ell D_e,\iota(J)}(\tau,z)$}
\label{compl-blowup}
\ee

There is however an important difference between \eqref{blowupf} and \eqref{relhch} for $\kappa=1$:
whereas the former is formulated in terms of the generating functions of stack invariants, the latter
is written directly in terms of the generating functions of VW invariants.
We recall that the reason for the use of stack invariants was that they are, in contrast to VW invariants,
well defined directly on walls of marginal stability, while,
as we will see below, a blow-up of $\IP^2$ leads to a polarization precisely corresponding
to one of such walls. Our results indicate that the generating functions $h^{\,S}_{N,\mu,J}$
and their completions $\whh^{\,S}_{N,\mu,J}$ are in fact also well defined on the walls!
In particular, the functions \eqref{genJhN} and \eqref{complFBJ} can be evaluated for any polarization
from the allowed two-parameter family $J\in\mbox{\rm Span}(c_1,\nv)^+$ and for any residue class $\mu$.
This is because the modular completions are actually smooth across the walls and provide
an unambiguous definition of the holomorphic generating functions everywhere including the walls.
In our case this is realized by a prescription for the contributions 
with vanishing arguments of sign functions in the kernel \eqref{kerg}.

We emphasize that this does not mean that the VW invariants are defined on the walls!
In fact, one can check (see the next section for an example) that the rational invariants extracted from $h^{\,S}_{N,\mu,J}$
on a wall do {\it not} lead to {\it integer} invariants
after application of the inverse of the formula \eqref{defcref}.
These are interesting questions whether these rational numbers have any geometrical meaning and
whether they can be converted to integer numbers.

\section{Generating functions for $\IP^2$ and their modular completion}
\label{sec-P2}

Now we are ready to turn to the main motivation of this work and to provide explicit expressions for
the generating functions of refined VW invariants of $\IP^2$ and their modular completions.
Using the construction of the previous section they can be related to the corresponding functions for the blow-up of $\IP^2$
which is known to be the Hirzebruch surface $\Fb_1$. Indeed, the bilinear form \eqref{extC} and the first Chern class $\chc_1$
are given in this case by
\be
\chC=\(\begin{array}{cc}
1 & 0 \\  0 & -1
\end{array}\),
\qquad
\chc_1=(3,-1)
\ee
and coincide with the intersection form and the first Chern class of $\Fb_1$ in the basis \eqref{baseF1}.
Thus, $D_2=[s]$ is the exceptional divisor $D_e$ of the blow-up, whereas $\iota(J)=D_1=[f]+[s]$.\footnote{In this basis
the null vector is $\nv(\Fb_1)=D_1-D_2$.}
As a result, we arrive at the following representation for the generating functions
\be
\fbox{$\displaystyle
h^{\IP^2}_{N,\mu}(\tau,z)=\frac{\Theta^{\Fb_1}_{N,\mu D_1+\ell D_2}(\tau,z;\{\Phi_n(D_1)\})}{B_{N,\ell}(\tau, z)}
$}
\label{hP2}
\ee

For $N=2$ and 3 we have checked that this expression does reproduce the generating functions computed in
\cite{Yoshioka:1994,Manschot:2010nc}. The integrability conditions \eqref{inegrcond}, which the generating functions of $\Fb_1$
should satisfy, are known to follow at $N=2$ from the periodicity property for the classical Appell function
and at $N=3$ from its generalization proven in \cite{Bringmann:2015}.
For higher ranks they remain unexplored.

The modular completion of the generating functions \eqref{hP2}, which was known until now only up to $N=3$,
can be given in the following two forms
\be
\fbox{$\displaystyle
\begin{split}
\whh^{\,\IP^2}_{N,\mu}(\tau,z)=&\,\frac{\Theta^{\Fb_1}_{N,\mu D_1+\ell D_2}(\tau,z;\{\whPhi_n(D_1)\})}{ B_{N,\ell}(\tau,z)}
\\
= &\, \frac{\eta(\tau)^N}{\prod_{j=1}^N \theta_1(\tau,(2j-1) z)}
\sum_{\ell=0}^{N-1}\theta_{N,\ell}(\tau,N z)\,\Theta^{\Fb_1}_{N,\mu D_1+\ell D_2}(\tau,z;\{\whPhi_n(D_1)\}
\end{split}$}
\label{whhP2}
\ee
Whereas the first representation is simpler to evaluate, the second provides a direct access to the modular properties
of $\whh^{\,\IP^2}_{N,\mu}$.

Finally, we note that the polarization $J=D_1$ for $\Fb_1$ is a wall of marginal stability for
(some of) the VW invariants with $\cmu=\cmu^2 D_2$.
Nevertheless, our version of the blow-up formula \eqref{hP2} perfectly works for all $\cmu$.
In the example below we demonstrate this for $N=3$ and $\mu=0$
as well as the fact mentioned in the end of the previous section
that a naive calculation of the refined VW invariants of $\Fb_1$ for this polarization
leads to rational numbers.

\subsubsection*{Example: $N=3$, $\mu=0$}

For these values of parameters the generating function of refined VW invariants of $\IP^2$ is known
and can be found, for instance, in \cite[Eq.(6.22))]{Manschot:2017xcr}. In our notations it reads
\bea
h^{\IP^2}_{3,0}(\tau,z) &= & -\frac{\I}{\theta_1(\tau,2z)^3\,\theta^{(3)}_{0}(\tau,z) }\[
\frac13\,\theta^{(3)}_{0}(\tau,z) +  y^4\sum_{k_1,k_2\in \IZ}
\frac{y^{-2k_1-4k_2}q^{k_1^2+k_2^2+k_1k_2+2k_1+k_2}}{(1-y^4q^{2k_1+k_2})(1-y^4q^{k_2-k_1})}
\right.
\nn\\
&& +\frac{2\I
  \eta(\tau)^3}{\theta_1(\tau,4z)}
  \( \sum_{k\in   \IZ}\frac{y^{-6k}q^{3k^2}}{1-y^6q^{3k}} -\frac12\, \theta_3(6\tau,6z) \)
\left.
-\frac{\eta(\tau)^6\,\theta_1(\tau,2z)}{\theta_1(\tau,4z)^2\,\theta_1(\tau,6z)}\].
\eea
Its first few terms of the expansion at small $q$ are given by
\be
h^{\IP^2}_{3,0}=\textstyle q^{-3/8}\[\frac{1}{3 (y^3-y^{-3})}
+\frac{y^{12}+2 y^{10}+4y^8 +\frac{16}{3}\,y^6+6y^4+6y^2 +\frac{19}{3}+6y^{-2}+6y^{-4}+\frac{16}{3}\,y^{-6}
+4y^{-8}+2 y^{-10}+y^{-12}}{y^3-y^{-3}}\,q^3
+O(q^4)\].
\label{exphP2N3}
\ee
This expansion allows to read off the rational invariants
$\bOm(\gamma,y)$ for small second Chern classes of the sheaf.
Inverting the formula \eqref{defcref}, one then extracts the integer valued invariants which encode
the Betti numbers of moduli spaces $\cM_{\gamma,J}$:
\be
h^{\IP^2}_{3,0}-\frac13\,h^{\IP^2}_{1,0}(3\tau,3z)=
\frac{q^{-3/8}}{y-y^{-1}}\Bigl[(y^2+1 + y^{-2}) (y^8 + y^4 +1+ y^{-4} + y^{-8})\, q^3 +O(q^4)\Bigr],
\label{inthP2N3}
\ee
in agreement with \cite[Eq.(A.39)]{Beaujard:2020sgs}.

Let us compare this result with \eqref{hP2}. The expansion of the numerator and denominator gives
\bea
h^{\Fb_1}_{3,0,D_1}(\tau,z)&=& q^{-1/2}\[\textstyle \frac{1}{3 (y^3-y^{-3})}
+\frac{y^2+1+y^{-2}}{3 (y-y^{-1})}\,q
+\frac{y^2+1+y^{-2}}{y-y^{-1}}\,q^2
\right.
\\
&+&\left.
\textstyle\frac{y^{12}+2 y^{10}+4y^8 +6y^6+9y^4+12y^2 +\frac{43}{3}+12y^{-2}+9y^{-4}+6y^{-6}+4y^{-8}+2 y^{-10}+y^{-12}}{y^3-y^{-3}}\,q^3
+O(q^4)\],
\nn
\\
B_{3,0}& =& q^{-1/8}\Bigl[1 + (y^2 +1+ y^{-2})^2 q + 3(y^2 +1+ y^{-2})^2 q^2
\Bigr.
\\
&&\Bigl.
+ \(2 y^6+ 9 y^4+ 18 y^2 +24 + 18y^{-2} + 9y^{-4} + 2y^{-6}\) q^3
+O(q^4) \Bigr].
\nn
\eea
It is immediate to check that the ratio of these two expansions indeed reproduces \eqref{exphP2N3}.
Thus, our blow-up formula \eqref{hP2} allows to bypass stack invariants even
when the polarization induced on the blow-up surface corresponds to a wall of marginal stability.
On the other hand, applying the same inverse formula as in \eqref{inthP2N3}, one obtains
\bea
h^{\Fb_1}_{3,0,D_1}(\tau,z)&-&\frac13\,h^{\Fb_1}_{1,0}(3\tau,3z)=\frac{q^{-1/2}}{y-y^{-1}}\Bigl[
\textstyle \frac{1}{3}\,(y^2+1+y^{-2})\,q
+(y^2+1+y^{-2})\,q^2
\\
&+&
\textstyle\(y^{-10}+y^{-8}+2y^{-6}+\frac{8}{3}\, y^{-4}+\frac{13}{3}\,y^{-2}+5+\frac{13}{3}\,y^2+\frac{8}{3}\,y^4
+2y^6+y^8+y^{10}\)q^3
+O(q^4)\Bigr].
\nn
\eea
Although this procedure does produce symmetric Poincar\'e polynomials as in \eqref{intOm}, their coefficients are not all integer.
Thus, the resulting numbers cannot be interpreted as Betti numbers of some moduli spaces and their interpretation remains open.

\section{Conclusions}
\label{sec-concl}

In this paper we studied generating functions of refined VW invariants
for $U(N)$ Vafa-Witten theory on a complex surface $S$
with $b_2^+(S)=1$ and $b_1(S)=0$. Such generating functions are higher depth
mock modular forms and possess non-holomorphic modular completions.
Our approach originates in the equation satisfied by these completions in the canonical chamber
of the moduli space \cite{Alexandrov:2019rth}
which, combined with constraints from the unrefined limit,
turns out be sufficiently restrictive to find the generating functions explicitly.
In this way, we obtained the following results, which all hold for arbitrary rank $N$:
\begin{itemize}
\item
Generalizing the results for Hirzebruch and del Pezzo surfaces found in \cite{Alexandrov:2020bwg} in the canonical chamber,
we extended them to the chambers corresponding to a two-parameter family of polarizations, which for Hirzebruch surfaces
cover the whole K\"ahler cone. In particular, in \eqref{genJhN} we provided explicit expressions for
the generating functions and in \eqref{complFBJ} for their completions.

\item
We also proposed the extension \eqref{exp-whhr} of the equation on the completion mentioned above to arbitrary polarization.

\item
We showed that this equation implies a version of the blow-up formula \eqref{relhch} for the generating functions of VW invariants,
which in contrast to the usual approach does not require the use of stack invariants.
A similar formula holds for the modular completions and it can nicely be rewritten in a manifestly modular way \eqref{compl-blowup}.

\item
Specializing to $S=\IP^2$, we arrived at the closed formula \eqref{hP2} for the generating functions $h^{\IP^2}_{N,\mu}$,
which have been computed before in a different form \cite{Manschot:2014cca},
and the new formula \eqref{whhP2} for their completions $\whh^{\,\IP^2}_{N,\mu}$.

\end{itemize}

Probably, the most unexpected result is that the generating functions \eqref{genJhN} which we proposed make sense
everywhere in the moduli space, including walls of marginal stability. This is implied by the fact
that they can be used in the blow-up formula and produce correct VW invariants on all surfaces related by
the blow-up or blow-down construction
(see Fig.1 in \cite{Beaujard:2020sgs} for a nice representation of such relations between different surfaces).
This fact was the key for avoiding stack invariants.
However, this leaves us with the open question of geometric interpretation of the numbers extracted from
the generating functions evaluated on a wall. We observed that inverting the formula \eqref{defcref},
one does get symmetric Poincar\'e polynomials, but their coefficients are not generically integer.
May be the first problem to consider is whether there is a general prescription which converts these rational numbers
into integer.

Finally, we would like to note that one of the motivations of this work was to develop methods which
might be useful in extending the approach based on the equation for the modular completion to compact CY threefolds,
the domain where it was originally derived. A serious complication arising for such extension is that the integer parameter $N$
is replaced by a vector $p^a$ valued in (dual to) the lattice of electric charges $\Lambda=H_2(\CY,\IZ)$.
A way to avoid this complication is to consider a CY with $b_2(\CY)=1$, the case analogous to $S=\IP^2$ for VW theory.
Thus, we expect that the results obtained here may find a direct application for the computation of DT invariants
of compact CYs with one K\"ahler modulus.

\section*{Acknowledgements}

The author is grateful to Jan Manschot and Boris Pioline for valuable discussions.

\appendix

\section{Theta series and their identities}
\label{ap-ident}

Let us recall the definition of vector valued Jacobi form of weight $w$ and index $m$.
This is a finite set of functions $\phi_\bfmu(\tau, z)$ with $\tau\in \IH$, $z\in \IC$ labelled by $\bfmu$
such that
\be
\begin{split}
\phi_{\bfmu}(\tau,z+k\tau+\ell)=&\, e^{- 2\pi\I m \( k^2\tau + 2 k z\)} \,\phi_{\bfmu}(\tau,z),
\\
\phi_{\bfmu}\(\frac{a\tau+b}{c\tau+d}, \frac{z}{c\tau+d}\)
=&\, (c\tau+d)^w \, e^{\frac{2\pi\I m c z^2}{c\tau+d}}\sum_\bfnu M_{\bfmu\bfnu}(\rho)\,\phi_{\bfnu}(\tau,z),
\end{split}
\label{Jacobi}
\ee
where $\rho=\scriptsize{\(\begin{array}{cc}
a & b \\ c & d
\end{array}\)}\in SL(2,\IZ)$ and we allowed for a non-trivial multiplier system $M_{\bfmu\bfnu}(\rho)$.

Known examples of (vector valued) Jacobi forms are theta series with a positive definite quadratic form.
In this work we encounter several such theta series.
Here we list their definitions and provide some useful identities which they satisfy.

\begin{itemize}

\item
Theta series
\be
\theta_{N,\ell}(\tau,z)=
\sum_{k\in N\IZ+\ell+\hf N} \q^{\frac{1}{2N}\, k^2}\, (-y)^k,
\ee
where $\ell=0,\dots ,N-1$, is a vector valued Jacobi form of weight 1/2 and index $N/2$.
For $N=1$ it reduces to the well known Jacobi theta function
\be
\theta_1(\tau,z)=\sum_{k\in \IZ+\hf}\q^{k^2/2} (-y)^k,
\label{deftheta1}
\ee
which is Jacobi form of weight 1/2 and index $1/2$.
It vanishes at $z=0$, whereas its first derivative gives
\be
\p_z\theta_1(\tau,0)=-2\pi \eta(\tau)^3,
\label{derth1}
\ee
where $\eta(\tau)=\q^{\frac{1}{24}}\prod_{n=1}^\infty(1-\q^n)$ is the Dedekind eta function,
modular form of weight 1/2.

\item
Theta function appearing in the blow-up formula
\be
\begin{split}
\theta^{(N)}_{\ell}(\tau,z)=&\, \sum_{\sum\limits_{i=1}^N k_i=\ell,\ k_i\in \IZ}
\q^{\frac{1}{2N}\sum_{i<j}(k_j-k_i)^2}\, y^{\sum_{i<j}(k_j-k_i)}
\\
=&\, \sum_{\sum_{i=1}^N a_i=0 \atop a_i\in\IZ+\frac{\ell}{N}}
\q^{- \sum_{i<j} a_i a_j}\, y^{\sum_{i<j}(a_j-a_i)}
\end{split}
\label{defthN}
\ee
transforms as vector valued Jacobi form of weight $\hf(N-1)$ and index $\frac16 (N^3-N)$.

\item
Theta series
\be
\theta^{(\vec N)}_{\ell,\vec\ell}(\tau,z)=\sum_{\sum_{i=1}^n k_i=\ell,\atop k_i\in N_i\IZ+\ell_i}
\q^{\sum_{i<j}\frac{(N_j k_i-N_i k_j)^2}{2NN_i N_j}}\, y^{\sum_{i=1}^n \ptt_i k_i},
\label{defthNll}
\ee
where vectors denote collections of $n$ components and
\be
\ptt_i=\sum_{j<i}N_j-\sum_{j>i}N_j\, ,
\label{defNv}
\ee
is a vector valued Jacobi form of weight $\frac{1}{2}\,n$ and index
$
\frac16\(N^3-\sum_{i=1}^n N_i^3\)$.

\end{itemize}
These theta series satisfy two important identities:
\bea
\sum_{\ell=0}^{N-1}\theta_{N,\ell}(\tau,\kappa' z)\,
\theta^{(N)}_{\ell}(\tau,\kappa z)&=&\prod_{j=1}^N \theta_1(\tau,\kappa_j z),
\qquad\qquad
\kappa_j=\kappa'+(2j-N-1) \kappa,
\label{diagThN}
\\
\theta^{(N)}_{\ell}(\tau,z)&=&\prod_{i=1}^n \[\sum_{\ell_i=0}^{N_i-1}\theta^{(N_i)}_{\ell_i}(\tau,z)\]
\theta^{(\vec N)}_{\ell,\vec\ell}(\tau,z),
\qquad
N=\sum_{i=1}^n N_i.
\label{decompThN}
\eea
They directly follow from the following lattice decompositions
\be
\begin{split}
\IZ^N=&\, \bcup\limits_{\ell=0}^{N-1} \(N\IZ+\ell\)\oplus\(\IZ^N/\IZ+\ell \ve_1\),
\\
\IZ^N/\IZ=&\,  \bcup\limits_{\ell_i=0}^{N_i-1}\[\bplus_{i=1}^n\(\IZ^{N_i}/\IZ+\ell_i \ve_1\)\]
\oplus \(\[\bplus_{i=1}^n\(N_i \IZ+\ell_i\)\]/\IZ\),
\end{split}
\ee
where factorization by $\IZ$ is the factorization by the diagonal group $\IZ \otimes (1,\dots,1)$,
whereas $\ve_1=(1,0,\dots,0)$. The necessity of shifts proportional to $\ell$ or $\ell_i$, 
which provide the so called glue vectors \cite{CSbook}, can be seen from the fact that a unimodular lattice $\IZ^N$ 
is decomposed into a direct sum of non-unimodular ones.

\section{Stack invariants and blow-up formula}
\label{ap-stack}

Stack invariants introduced in \cite{joyce2008configurations} are naturally organized in the generating functions
similar to \eqref{defhVWref}
\be
\label{defHstack}
H^S_{N,\mu,J}(\tau,z) =
\sum_{n\geq 0}
\cI_J(\gamma,y)\,
\q^{N\(\Delta(F) - \tfrac{\chi(S)}{24}\)}.
\ee
We do not give here their precise definition, but just note that they are defined using
the so-called slope stability (also known as $\mu$-stability) condition, in contrast to VW invariants for which
Gieseker stability is relevant, and the relation between the two types of invariants is given by
\be
h^S_{N,\mu,J}=\sum_{\sum_{i=1}^k(N_i,\mu_i)=(N,\mu)\atop \nu_J(N_i,\mu_i)=\nu_J(N,\mu)} \frac{(-1)^k}{k}
\prod_{i=1}^k H^S_{N_i,\mu_i,J}\, ,
\ee
with inverse relation
\be
H^S_{N,\mu,J}=\sum_{\sum_{i=1}^k(N_i,\mu_i)=(N,\mu)\atop \nu_J(N_i,\mu_i)=\nu_J(N,\mu)} \frac{1}{k!}
\prod_{i=1}^k h^S_{N_i,\mu_i,J}\, .
\ee
Here
\be
\nu_J(N,\mu)=\frac{\mu\cdot J}{N}
\ee
is the slope defining the stability condition.
Note that $\cI_J(\gamma,y)$ have higher order poles at $y=\pm 1$, and possess
more complicated symmetry and integrality properties than $\Omega_J(\gamma,y)$. On the other hand, their
behavior under wall crossing is simpler and, in contrast to DT or VW invariants, they can be defined
directly on walls of marginal stability.

The latter property makes them particularly useful in the formulation of the blow-up formula
which relates the generating functions of topological invariants of a surface $S$ and of its blow-up $\pi:\chS\to S$
\cite{Yoshioka:1996,0961.14022,Li:1998nv}. Although it can be formulated for the generating functions of VW invariants
$h^S_{N,\mu,J}$, in that case it is restricted to the parameters satisfying $\gcd(N,\mu\cdot J)=1$.
The reason of this restriction is that otherwise one sits on a wall of marginal stability.
Instead, the stack invariants allow to avoid this restriction and their generating functions on $S$ and $\chS$
are related by
\be
H^{\chS}_{N,\pi_\star(\mu)+\ell D_e,\pi_\star(J)}(\tau,z)=B_{N,\ell}(\tau,z)\, H^S_{N,\mu,J}(\tau,z),
\label{blowupf}
\ee
where $D_e$ is the exceptional divisor of the blow-up $\pi$, and
\be
\label{defBNk}
B_{N,\ell}(\tau,z) = \frac{\theta^{(N)}_{\ell}(\tau,z)}{\eta(\tau)^N}.
\ee
Here $\eta(\tau)$ is the Dedekind function and $\theta^{(N)}_{\ell}$ is the theta series \eqref{defthN},
so that $B_{N,\ell}$ is a vector valued Jacobi form of weight $-\hf$ and index $\frac16 (N^3-N)$ which we call `blow-up function'.

Finally, we note that for $S=\Fb_m$ and $\Bb_m$ and for the polarization $J=\nv$,
determined by the same null vector \eqref{nullv}
which enters the construction of $h_{N,\mu}$,
the generating functions of stack invariants take particularly simple form.
They read \cite{Manschot:2011ym,Mozgovoy:2013zqx}
\be
H^S_{N,\mu}:=H^S_{N,\mu,\nv}=\delta^{(N)}_{\nv\cdot\mu} \, H_N \prod_{\alpha=3}^{b_2(S)}B_{N,\mu^\alpha},
\qquad
H_N=\frac{\I (-1)^{N-1} \eta(\tau)^{2N-3}}
{\theta_1(\tau,2Nz)\, \prod_{m=1}^{N-1} \theta_1(\tau,2mz)^2}\, ,
\label{gfHN}
\ee
where
\be
\delta^{(n)}_x=\left\{ \begin{array}{l} 1 \quad \mbox{if } x=0 \mbox{ mod }n, \\ 0 \quad \mbox{otherwise,} \end{array} \right.
\label{defdelta}
\ee
$\theta_1(\tau,z)$ is the Jacobi theta function \eqref{deftheta1},
and in the last factor, which is relevant only for del Pezzo surfaces, $\mu^\alpha$, $\alpha\ge 3$,
are the components of the residue class $\mu$ along $m-1$ exceptional divisors of $\Bb_m$, i.e. $\mu=\mu^\alpha D_\alpha$
where $D_\alpha$ is the same basis as in \eqref{dataBk}.
Clearly, this factor is a direct consequence of the blow-up formula \eqref{blowupf}.

\section{Generalized error functions}
\label{ap-generr}

Theta series with an indefinite quadratic form are potentially divergent.
To ensure convergence their kernel must be a combination of sign functions which restrict
the sum over lattice to a sublattice where the quadratic form is positive definite.
An example of such kernel is \eqref{kerg}. However, the insertion of sign functions spoil their modular properties
and make them examples of (higher depth) mock modular forms.
Nevertheless, there is a simple recipe to construct their modular completion \cite{Zwegers-thesis,Alexandrov:2016enp,Nazaroglu:2016lmr}.

To this end, let us define the generalized error functions
introduced in \cite{Alexandrov:2016enp,Nazaroglu:2016lmr}
\bea
E_n(\cM;\vu)&=& \int_{\IR^n} \de \vu' \, e^{-\pi(\vu-\vu')^{\rm tr}(\vu-\vu')} 
\prod_{i=1}^n \sgn(\cM^{\rm tr} \vu')_i\, ,
\label{generr-E}
\eea
where $\vu=(u_1,\dots,u_n)$ is $n$-dimensional vector and $\cM$ is $n\times n$ matrix of parameters.
The detailed properties of these functions can be found in \cite{Nazaroglu:2016lmr}.
This is however not enough since, to provide a kernel of theta series which is defined over a $d$-dimensional
lattice $\Lat$ with a bilinear form $\bfx\bpt\bfy$ of signature $(d-n,n)$, 
we need a function depending on a $d$-dimensional vector.
Such functions, called boosted generalized error functions, are defined by
\be
\Phi_n^E(\{\bfv_i\};\bfx)=E_n(\{\bfb_i\bpt\,\bfv_j\} ;\{\bfb_i\bpt\, \bfx\}),
\label{generrPhiME}
\ee
where the vectors $\bfv_i$ are supposed to span a positive definite subspace and
$\bfb_i$ form an orthonormal basis in this subspace.
It can be shown that $\Phi_n^E$ does not depend on the choice of this basis
and at large $\bfx$ reduces to $\prod_{i=1}^n \sgn (\bfv_i\bpt\,\bfx)$.
Furthermore, it solves certain differential equation which ensures modularity of the corresponding theta series.

Thus, to construct the modular completion of a theta series whose kernel is a combination of sign functions,
one can apply the following recipe. Let the dependence on the elliptic parameter
is captured by the factor $y^{\bfptt\bpt\bfq}$
where $\bfq\in \Lat$ is the summation variable. Then each term in the kernel of the form $\prod_{i=1}^n \sgn (\bfv_i\bpt\,\bfq+\psi_i)$
must be replaced by $\Phi_n^E(\{\bfv_i\};\bfx)$ where
\be
\bfx=\sqrt{2\tau_2}(\bfq+\beta\bfptt)
\label{vecx}
\ee
and $\beta=-\frac{\Im z}{2\tau_2}$ is the imaginary part of the elliptic parameter.
The necessity of the $\beta$-dependent shift can be seen, for example, from the periodicity
condition which any Jacobi form must satisfy.
It is important that if one of the vectors $\bfv_i$ is null, it reduces the rank of the generalized error function.
Namely, for $\bfv_\ell^2=0$, one has
\be
\Phi_n^E(\{\bfv_i\};\bfx)=\sgn (\bfv_\ell\,\bpt\,\bfx)\,\Phi_{n-1}^E(\{\bfv_i\}_{i\in \Zv_{n}\setminus\{\ell\}};\bfx).
\label{Phinull}
\ee
In other words, for such vectors the completion is not required.

\section{Coefficient $\Rv_n$}
\label{ap-Rn}

In this appendix we provide the definition of the coefficient $\Rv_n$ appearing
in the formula for the modular completion \eqref{exp-whhr}. It carries a non-holomorphic dependence
on both $\tau=\tau_1+\I\tau_2$ and $z=\alpha-\tau\beta$ and thus makes the completion also non-holomorphic.

The formula for $\Rv_n$ reads as follows \cite{Alexandrov:2019rth}
\be
\Rv_n(\{\gama_i\},J;\tau_2,\beta)= \Sym\left\{\sum_{T\in\IT_n^{\rm S}}(-1)^{n_T-1}
\sEp_{v_0}\prod_{v\in V_T\setminus{\{v_0\}}}\sEf_{v}\right\},
\label{solRnr}
\ee
where $\Sym$ denotes symmetrization (with weight $1/n!$) with respect to charges $\gama_i$,
the sum goes over so-called Schr\"oder trees with $n$ leaves (see Figure \ref{fig-Rtree}), i.e. rooted planar
trees such that all vertices $v\in V_T$ (the set of vertices of $T$ excluding the leaves) have $k_v\geq 2$ children,
$n_T$ is the number of elements in $V_T$, and $v_0$ labels the root vertex.
The vertices of $T$ are labelled by charges so that the leaves carry charges $\gama_i$, whereas the charges assigned to other vertices
are given recursively by
the sum of charges of their children, $\gama_v\in\sum_{v'\in\Ch(v)}\gama_{v'}$.
Finally, to define the functions $\sEf_v$ and $\sEp_v$, let us consider
a set of functions $\sE_n$ depending on $n$ charges, $\tau_2$ and $\beta$, whose explicit expressions will be given shortly.
Given this set, we take
\be
\begin{split}
\sEf_n(\{\gama_i\})=&\,\under{\lim}{\tau_2\to\infty}\sE_n\(\{\gama_i\},\tau_2, -\frac{\Im z}{\tau_2}\),
\\
\sEp_n(\{\gama_i\},\tau_2,\beta)=&\,\sE_n(\{\gama_i\},\tau_2,\beta)-\sEf_n(\{\gama_i\}),
\end{split}
\label{redef-cErf}
\ee
so that $\sEf_n$ does not depend on $\tau_2$ (and $\beta$),
whereas the second term $\sEp_n$ turns out to be exponentially suppressed as $\tau_2\to\infty$ keeping
the charges $\gama_i$ fixed. Then, given a Schr\"oder tree $T$,
we set $\sE_{v}\equiv \sE_{k_v}(\{\gama_{v'}\})$ (and similarly for $\sEf_{v}, \sEp_{v}$)
where $v'\in \Ch(v)$ runs over the $k_v$ children of the vertex $v$.

\lfig{An example of Schr\"oder tree contributing to $\Rv_8$. Near each vertex we showed the corresponding factor
using the shorthand notation $\gamma_{i+j}=\gamma_i+\gamma_j$.}
{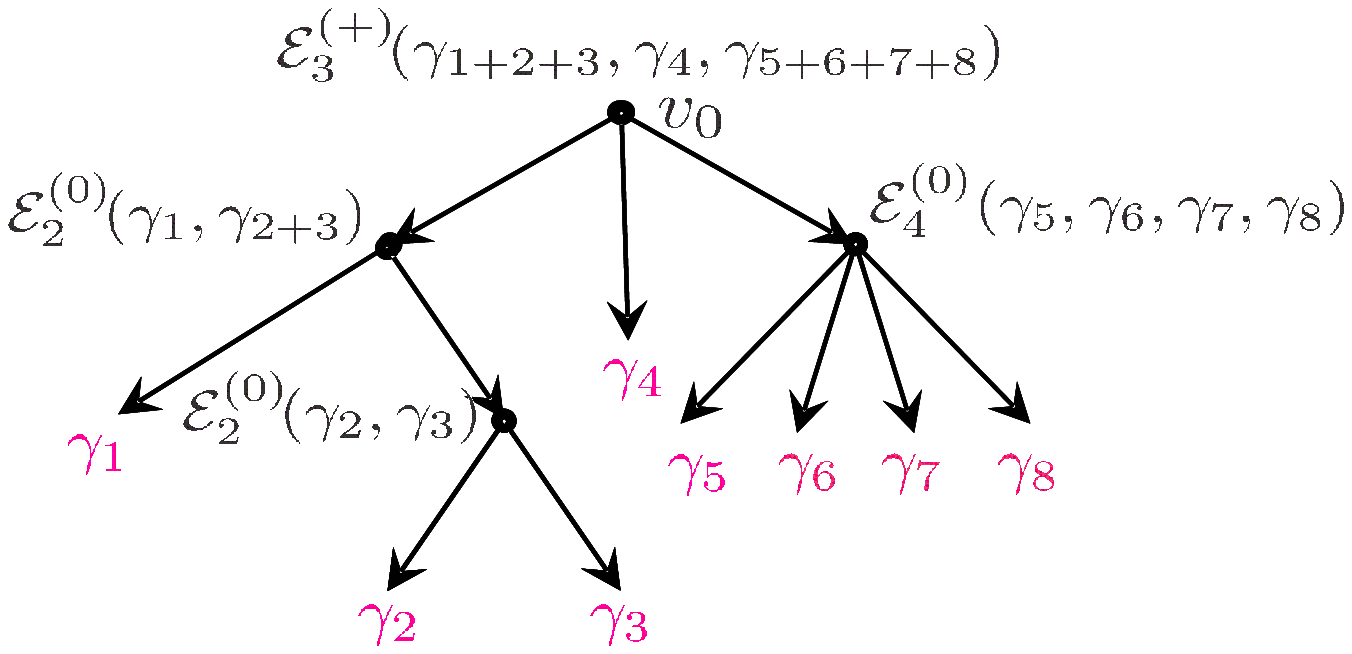}{9.5cm}{fig-Rtree}{-1.2cm}

It remains to provide the functions $\sE_n$. They are given by
\be
\sE_n(\{\gama_i\},\tau_2,\beta)= \Phi^E_{n-1}(\{ \bfv_{\ell}(J)\};\bfx),
\label{Erefsim}
\ee
where $\Phi^E_{n-1}$ are (boosted) generalized error functions described in appendix \ref{ap-generr},
which depend on $nb_2$-dimensional vectors with the following components
\be
\begin{split}
\bfx_i^\alpha=&\, \sqrt{2\tau_2}\(\tfrac{1}{N_i}\,C^{\alpha\beta}q_{i,\beta}+\beta \ptt_i c_1^\alpha\),
\qquad i=1,\dots, n,
\\
\bfv_{\ell,i}^\alpha(J) =&\, \(\sN_\ell\delta_{i>\ell}-(N-\sN_\ell)\delta_{i\le \ell}\)J^\alpha,
\qquad \ell=1,\dots, n-1,
\end{split}
\label{vectors}
\ee
where $\sN_\ell=\sum_{k=1}^\ell N_k$ and $\ptt_i=\sum_{j<i}N_j-\sum_{j>i}N_j$.

The meaning of the vectors \eqref{vectors} can be understood as follows.
The first vector $\bfx$ simply combines all charges $q_i$ into one vector and has the form \eqref{vecx} required
by modularity. This follows from the identity
\be
\sum_{i<j}\gamma_{ij}(c_1)=c_1\cdot\sum\limits_{i=1}^n \ptt_i q_i=\bfptt\bpt\bfq,
\label{sumgam}
\ee
where we introduced the vectors
\be
\bfptt_{i}^\alpha = \ptt_i c_1^\alpha,
\qquad
\bfq_i^\alpha= \tfrac{1}{N_i}\,C^{\alpha\beta}q_{i,\beta}
\label{vectors2}
\ee
and the bilinear form\footnote{We use different multiplication symbols to distinguish between
bilinear forms on different spaces: $\cdot$ denotes contraction of $b_2$-dimensional vectors using $C_{\alpha\beta}$
(or its inverse), whereas $\bpt$ is used for $nb_2$-dimensional vectors.}
\be
\bfx\bpt\bfy=\sum_{i=1}^n N_i \, x_i\cdot y_i.
\label{biform}
\ee
The second vector is designed so that
\be
\bfv_k(J)\,\bpt\, \bfx =\sqrt{2\tau_2}\,\Bigl(\Gamma_k(J)+\beta N\sN_k (N-\sN_k)\, J\cdot c_1\Bigr).
\label{defGam}
\ee
The first term is nothing but the quantity introduced in \eqref{defbk} and appearing as the argument of
one set of sign functions in the kernel \eqref{kerg} defining the generating functions.
Note also that the second set of sign functions in \eqref{kerg} can be obtained in a similar way, namely
\be
\bfw_{k,k+1}\bpt\,\bfx=\sqrt{2\tau_2}\,\Bv_k,
\qquad
\bfw_{k\ell,i}^\alpha= \(N_k\delta_{i\ell}-N_\ell\delta_{ik}\) \nv^\alpha.
\ee
The difference however is that in the first set the $\beta$-dependent shift present in \eqref{defGam}
disappears. This is because this set as well as the special treatment of vanishing arguments originates in $\sEf_n$.
Indeed, evaluating the limit in \eqref{redef-cErf}, one reproduces the $\nv$-independent term in the kernel \eqref{kerg} 
\cite{Alexandrov:2019rth}:
\be
\sEf_n(\{\gama_i\})=\sum_{\cJ\subseteq\Zv_{n-1}} e_{|\cJ|}\,\delta_\cJ(v)
\prod_{k\in \Zv_{n-1}\setminus \cJ}\sgn(\Gamma_k(J))\, .
\label{rel-gEf-zero}
\ee

\providecommand{\href}[2]{#2}\begingroup\raggedright\endgroup


\end{document}